\documentclass[apj]{emulateapj}
\usepackage{amsmath}

\begin{document}

\title{GALAXY-MASS CORRELATIONS ON 10 MPC SCALES IN THE DEEP LENS SURVEY}

\author{A. Choi\altaffilmark{1,2,4}, J. A. Tyson\altaffilmark{1}, C. B. Morrison\altaffilmark{1}, M. J. Jee\altaffilmark{1}, S. J. Schmidt\altaffilmark{1}, V. E. Margoniner\altaffilmark{3}, D. M. Wittman\altaffilmark{1}}

\altaffiltext{1}{Physics Department, University of California, One Shields Avenue, Davis, CA 95616}
\altaffiltext{2}{Scottish Universities Physics Alliance, Institute for Astronomy, University of Edinburgh, Royal Observatory, Blackford Hill, Edinburgh EH9 3HJ}
\altaffiltext{3}{Physics and Astronomy Department, California State University, Sacramento, CA 95819}
\altaffiltext{4}{choi@roe.ac.uk}

\begin{abstract}

We examine the projected correlation of galaxies with mass from small scales ($<$few hundred kpc) where individual dark matter halos dominate, out to 15 Mpc where correlated large-scale structure dominates. We investigate these profiles as a function of galaxy luminosity and redshift.  Selecting 0.8 million galaxies in the Deep Lens Survey, we use photometric redshifts and stacked weak gravitational lensing shear tomography out to radial scales of 1 degree from the centers of foreground galaxies.
We detect correlated mass density from multiple halos and large-scale structure at radii larger than the virial radius, and find the first observational evidence for growth in the galaxy-mass correlation on 10 Mpc scales with decreasing redshift and fixed range of luminosity.
For a fixed range of redshift, we find a scaling of projected halo mass with rest-frame luminosity similar to previous studies at lower redshift.
We control systematic errors in shape measurement and photometric redshift, enforce volume completeness through absolute magnitude cuts, and explore residual sample selection effects via simulations.

\end{abstract}


\section{Introduction}

The presence of dark matter in the Universe is well-established and contributes significantly to structures ranging from galaxies to superclusters. Less understood is the distribution and evolution of dark matter correlated with galaxies over larger volumes beyond the galaxy virial radius.  The WMAP cosmic microwave background (CMB) result for $\Omega_{\rm m} = 0.27\pm 0.03$ \citep{2011ApJS..192...18K} is over twice that derived from N-body model fits to lensing studies of the inner mass profile of luminous galaxies together with cosmic luminosity density \citep{2012ApJ...746...38M}.  This is not surprising, since about half the dark matter (DM) is expected to be in virialized halos \citep{2006ApJ...639..590F}.  Detailed measurements of the mass distribution on large scales around galaxies, or ``galaxy-mass correlation,'' as a function of galaxy properties can thus be a diagnostic of structure formation and evolution.  A universal mechanism for hierarchical structure formation was developed by \citet{1974ApJ...187..425P}, drawing on bottom-up structure formation ideas of \citet{1965ApJ...142.1317P} and \citet{1968MNRAS.141....1S}. More generally, growth of very large scale mass structures is cosmology dependent, and is one of the probes of the physics of dark energy \citep{2006astro.ph..9591A}.

Weak gravitational lensing (WL) is the only direct probe that can both measure the mass profile associated with galaxies over a wide range of radii crossing the virialization and turnaround scales and does not require assumptions about the dynamical state or baryon content of the system in question.  WL is an inherently statistical technique: many galaxies are required. The dimensionless shear signal ranges from less than 0.1\% (large-scale structure) to 1\% (galaxies) to 10\% (clusters of galaxies).  The WL signal from foreground DM halos lensing more distant galaxies is commonly known as ``galaxy-galaxy lensing'' (GGL) and is measured by cross-correlating the positions of foreground DM halos (as traced by their resident galaxies) with the lensing-induced shear of background galaxies.
Most GGL results in the literature focus on the inner parts of the DM halo mass profile inside the virial radius, which is dominated by contributions from individual halos.  Observations of luminosity scaling of the halo mass have been reported by \citet{2005ApJ...635...73H}, \citet{2006MNRAS.368..715M}, \citet{2006A&A...455..441K}, and \citet{2011arXiv1107.4093V}.  Recent studies have also tackled the problem of redshift evolution of halo mass \citep{2010ApJ...716.1579L,2012ApJ...744..159L}.  However, the study of GGL outside the virial radius is difficult from both observational and theoretical standpoints.  In observations, residual systematics can masquerade as the lensing signal on large scales \citep{2005MNRAS.361.1287M}.  On the theory side, we need detailed N-body simulations and ray-tracing studies or comprehensive statistical methods to interpret the signal \citep[e.g.][]{2005MNRAS.362.1451M,2008MNRAS.388....2H,2012ApJ...746...38M}.

With five 4 square degree deep fields imaged in BVRz$^{\prime}$, the Deep Lens Survey offers the possibility of measuring galaxy-mass correlations tomographically over a wide range of projected radii well beyond the halo virial radius, and over a useful range of redshift.  In this paper, we investigate how the average galaxy-mass correlation over a wide range of radii varies with luminosity and redshift up to $z$=0.75.  We show the first observational evidence for growth of projected galaxy-mass correlations on 10 Mpc scales.  Section 2 briefly describes the formalism for weak lensing; Section 3 describes the data with particular attention to the measurement of both image shapes and photometric redshifts; Section 4 describes our investigation of photometric redshift error propagation; we present the results in Section 5 and compare them with previous observations and simulations; and we discuss and summarize our results in Section 6.  The Appendix contains further details of systematics tests and correlation matrices.  Throughout, we assume a $\Lambda CDM$ universe with $H=70$ km Mpc$^{-1} $s$^{-1}$, $\Omega_{M}=0.27$, and $\Omega_{\Lambda}=0.73$.  Distances are given in units of comoving Mpc.

\section{Formalism}

\subsection{Gravitational Lensing}
The projected mass overdensity of a lens is related to the induced shear of source galaxies through
\begin{equation}
\gamma_{T}\Sigma_{\rm crit} = \bar{\Sigma}(<R)-\bar{\Sigma}(R) ~~ \equiv \Delta\Sigma
\end{equation}
where R is the transverse separation between lens and source on the sky, and $\gamma_{T}$ is the tangential shear azimuthally averaged over an annulus with radius R centered at the lens. $\Delta\Sigma$ is the mean excess projected surface density, which is defined as the difference between the average surface density within radius R, $\bar{\Sigma}(<R)$, and the projected surface density at that radius $\bar{\Sigma}(R)$ \citep{1991ApJ...370....1M}.  In this paper, we refer to $\Delta \Sigma$ as a probe of galaxy-mass correlations, not to be confused with the 3D galaxy-mass correlation function $\xi_{gm}$.  $\Delta \Sigma$ and $\xi_{gm}$ are related in the following way:
\begin{eqnarray}
\Sigma(R) &=& \int \bar{\rho} \xi_{gm}(\sqrt{R^{2}+\chi^{2}})d\chi \\
\bar{\Sigma} &=& \frac{2}{R^{2}}\int_{0}^{R} \Sigma(R^{\prime})R^{\prime}dR^{\prime}
\end{eqnarray}
$\bar{\rho}$ is the mean matter density, and $\chi$ is the line of sight distance.
The critical surface density $\Sigma_{\rm crit}$ is defined as:
\begin{equation}\label{eq:sigcriteq}
\Sigma_{\rm crit} \equiv \frac{c^{2}}{4\pi G}\frac{D_{S}}{D_{L}D_{LS}(1+z_{L})^{2}}
\end{equation}
where D$_{L}$, D$_{S}$, and D$_{LS}$ are the angular diameter distances to the lens, source, and between lens and source.  We will notate lens and source redshifts as z$_{L}$ and z$_{S}$, respectively.  The factor of $(1+z_{L})^{2}$ arises from our use of comoving coordinates.  This choice is motivated by the relatively long baseline in lens redshift and radial separation scales that we investigate in this paper.

If we correct for the systematic effects introduced in the observing process and assume that gravitational lensing is the only phenomenon systematically distorting the shapes of the observed background galaxies around the position of foreground galaxies in the weak limit where both shear $\gamma$ and convergence $\kappa=\Sigma/\Sigma_{\rm crit}$ are much smaller than 1, $\gamma_{T}$ is related to the observed ellipticity $e_{T}$ via:
\begin{equation}\label{eq:obsell}
e_{T} = \gamma_{T}+e_{T}^{\rm int}
\end{equation}
where e$_{T}$ is the ellipticity component perpendicular to the line transversely connecting the lens and source and e$_{T}^{int}$ is the intrinsic ellipticity.  Note that the definition of ellipticity used here is $e=(a-b)/(a+b)$ where $a$ and $b$ are the semimajor and semiminor axes, respectively. When averaging over an ensemble of randomly oriented source galaxies, the $\langle$e$_{T}^{int}\rangle$ term drops out, leaving:
\begin{equation}\label{eq:sh_def}
\langle \gamma_{T} \rangle = \langle e_{T} \rangle
\end{equation}
While Equation~\ref{eq:obsell} is only valid in the limit e$_{T}^{int}\ll 1$, we account for this by normalizing to simulations as described in Section~\ref{shapes}.

Our estimator for the differential surface mass density for stacked lens galaxies is a lensing signal-to-noise variance weighted average summed over lens-source pairs \citep{2004AJ....127.2544S,2008MNRAS.386..781M}:
\begin{eqnarray}\label{eq:dsig_calc}
\widetilde{\Delta \Sigma} &=& \frac{\sum_{i}^{N_{\rm lens}}\sum_{j}^{N_{\rm src}}\tilde{w}_{j,i} \gamma_{T}^{j,i}\tilde{\Sigma}_{{\rm crit},j,i}}{\sum_{i}^{N_{\rm lens}}\sum_{j}^{N_{\rm src}}\tilde{w}_{j,i}} \\
\tilde{w}_{j,i} &=& \frac{1}{\tilde{\Sigma}_{{\rm crit},j,i}^{2}(\sigma_{\rm SN}^{2}+\sigma_{e}^{2})}
\end{eqnarray}
where weights $\widetilde{w_{j,i}}$ depend on angular diameter distances (see Equation~\ref{eq:sigcriteq}), the shape noise $\sigma_\textrm{SN}$, and the ellipticity measurement error $\sigma_e$. The shape noise is the rms ellipticity per component for the source sample. In practice, we measure $\widetilde{\Delta \Sigma}$ for each lens-source pair and bin the values by the projected comoving separation between the lens and source. The relationship between the estimated differential surface mass density $\widetilde{\Delta \Sigma}$ and the true quantity $\Delta \Sigma$ is investigated using simulations in Section~\ref{photozsim}, and for the rest of the paper the tilde is dropped.

\subsection{Halo Mass Profiles}

The mass density profiles of dark matter halos are commonly modeled using Navarro-Frenk-White (NFW) functions, which were motivated by dark matter simulations \citep{1996ApJ...462..563N}.
\begin{equation}\label{eq:density_nfw}
  \rho_{\rm NFW}(r) = \frac{\bar{\delta}\rho_{c}(z)}{(\frac{r}{r_{s}})(1+\frac{r}{r_{s}})^{2}}
\end{equation}
where $$\bar{\delta}=\frac{200 c^{3}}{3(\ln (1+c)+\frac{c}{1+c})}$$ is a characteristic density,  $\rho_{c}(z)$ is the critical density at redshift z, and $r_{s}$ is the mass scale radius. The concentration $c={r_{200}}/{r_{s}}$, and $r_{200}$ is the radius at which the mean enclosed mass density of the halo equals $200 \times \rho_{c}$.  M$_{200}$ is the mass enclosed within $r_{200}$, and we use the terms ``M$_{200}$'' and ``halo mass'' (M$_{\rm halo}$) interchangeably, and we note the specific cases where M$_{\rm halo}$ means something other than M$_{200}$.

We obtain the surface mass density profile $\Sigma_{\rm NFW} (R)$ by integrating the three-dimensional density over the line of sight. A second integration yields the mean surface mass density inside the projected radius:   $\Sigma_{\rm NFW}(\leq R)$. The expressions for these quantities can be found in \citet{1996A&A...313..697B,2000ApJ...534...34W,2003MNRAS.339..387Y}.

The enclosed mass, radial scale, and concentration are degenerate.  To reduce the degeneracy, we fix the concentration using the relation found in \citet{2008MNRAS.390L..64D} based on N-body simulations using the WMAP 5 cosmology \citep{2009ApJS..180..330K}, effectively turning the NFW model into a single parameter profile.

\begin{equation}
  c(M_{\rm 200},z) = 5.71(\frac{M_{\rm 200}}{2\times10^{12}h^{-1}M_{\odot}})^{-0.084}(1+z)^{-0.47}
\end{equation}

In this work, we do not fit a full halo model, which can be used to isolate contributions from individual halos, satellite halos, and neighboring halos as is done in, e.g. \citet{2005MNRAS.362.1451M} and \citet{2011arXiv1107.4093V}.  As it is a useful tool for interpretation purposes, we investigate the halo model as a fitting function in future work and simply comment on how its use might impact our results in this work.

\section{Data}

The Deep Lens Survey (DLS) consists of $\sim$100 nights of BVRz$^{\prime}$ imaging in five widely separated fields DLS F1-F5, each subtending 4 square degrees of the sky \citep{2002SPIE.4836...73W}.  Each field is composed of nine ``subfields,'' each subtending an area (40$^{\prime}$ x 40$^{\prime}$) slightly larger than the camera field of view and covered with dithers of $\sim$200''.  F1 and F2 were observed by Mosaic-1 \citep{1998SPIE.3355..577M} at the NOAO/KPNO 4-m Mayall Telescope, and F3-F5 were observed by Mosaic-2 at the NOAO/CTIO 4-m Blanco Telescope.  The DLS was designed to go deep enough at high S/N to yield good photometric redshifts and to measure galaxy shapes at low surface brightness, while being wide enough to average over sample variance and to contain enough galaxies to minimize shot noise in precision studies of mass.  The observing strategy was tuned specifically for lensing studies: imaging in the R band only when the seeing was 0.9'' or better. The effective exposure time in the R band is 18,000s, and the effective exposure time in the other three bands is 12,000s. We performed internal photometric calibration using the $\ddot{u}bercal$ method \citep[][]{2008ApJ...674.1217P,2011arXiv1111.2058W} and external calibration using \citet{1992AJ....104..340L} standard stars and SDSS \citep{2000AJ....120.1579Y,2011arXiv1111.2058W}. Details of the photometric stacking procedure can be found in the DLS technical summary paper (Wittman et al., in prep).  We created object catalogs using SExtractor in dual-image mode using the R-band image for detections \citep{1996A&AS..117..393B}. All magnitudes in this paper are SExtractor MAG$\_$AUTO (henceforth referred to as $m_{R}$) calibrated to the Vega system and have been corrected for galactic extinction using the reddening maps of \citet{1998ApJ...500..525S} and for point spread function (PSF) variation using the ColorPro algorithm of \citet{2006AJ....132..926C}.  To convert the R band magnitude to the AB system, one must add 0.20.\footnote[1]{See http://dls.physics.ucdavis.edu/calib/vegaab.html}  The survey is 50\% complete for object recovery to m$_{R}=26$, m$_{B,V}=25.5$, and m$_{z}=24.5$ where the subscripts B, V, and z refer to the B, V, and z bands, respectively.  After conservative quality and S/N cuts described below, we utilize a subset of $\sim$800,000 galaxies.  The analysis processes for shear and photometric redshift measurement are described in Sections~\ref{shapes} and \ref{photozsec} below.

\subsection{\label{shapes}Shapes}

In order to correct the observed ellipticities to the intrinsic ellipticities in Equation \ref{eq:sh_def}, we must account for contributions from instrumental and atmospheric effects.  As in \citet{2007PASP..119.1403J}, we use an interpolation of principal components characterizing star ellipticities and sizes measured on individual exposures to produce a PSF model for the co-add image.  To summarize the procedure, we select high S/N isolated stars using an iterative algorithm based on half-light radius vs magnitude, measure principal components (eigen-PSF) and coefficients (amplitude along the eigen-PSF), fit 3rd order polynomials to the spatial variation of the coefficients, and generate an effective PSF model after interpolating or stacking the PSF models for the individual exposures.  Twenty principal components or eigen-PSFs per exposure can robustly reproduce the observed variation of the PSF ellipticity and size ($<$99\% of the variance) within each CCD.   We measure galaxy semi-major and semi-minor axes in the co-add image by fitting elliptical Gaussians convolved with the PSF model using methodology as in \citet{2002AJ....123..583B}.  Galaxies for which this fit converges poorly are not used in the output catalog of ellipticities. This {\it Stack-Fit} measurement method has been tested on simulations of the Large Synoptic Survey Telescope \citep[LSST;][]{2011PASP..123..596J}, and full details are given in \citep{JeeShear2012}.  The difference from the procedure described in \citep{2011PASP..123..596J} is that the images are matched to the DLS:  the simulated shear maps are convolved with spatially varying PSFs similar to what is found in the DLS , down-sampled to the DLS pixel scale, and have noise added to match the DLS depth.  In summary, since the derived shear is increasingly underestimated with increasing noise, a S/N or magnitude-dependent calibration factor is necessary to recover the true shear from the measured shear.  The true shear $\gamma^{\rm true}$ can be related to the observed shear $\gamma^{\rm obs}$ using both a multiplicative component, $m_{\gamma}$ and an additive component C via $\gamma^{\rm true} = m_{\gamma}\gamma^{\rm obs} + $C.  Based on the above image simulations, the multiplicative calibration factor $m_{\gamma}$ can be parameterized by:
\begin{equation}\label{eq:shcalib}
m_{\gamma} = 6 \times 10^{-4}(m_{R}-20)^{3.26}+1.036
\end{equation}
The additive calibration factor C is negligible, as it is 10\% of the statistical error on all radial scales relevant to this work.  In this analysis, we use source galaxies with 22$<$m$_{R}$$<$24.5, corresponding to 1.04$<$$m_{\gamma}$$<$1.12.  Additionally, we require that the semi-minor axis be greater than 0.4 pixels to eliminate very small galaxies that might suffer from pixellation effects, $\sigma_{e}<0.3$, and STATUS=1, where STATUS is the convergence indicator for the minimization routine used for fitting the elliptical Gaussians, MPFIT \citep{2009ASPC..411..251M}.  For our data, shape shot noise $\sigma_\textrm{SN} \sim 0.3$ (see Equation~\ref{eq:dsig_calc}).

Given the position of a foreground lens, the tangential and 45$^{\circ}$-rotated ellipticity components are calculated from the semi-major and semi-minor ellipticity components e$_{1}$ and e$_{2}$ as follows:
\begin{eqnarray}
e_{t} &=& -e_{1}\cos 2\theta - e_{2}\sin 2\theta \\
e_{x} &=& e_{1}\sin 2\theta - e_{2}\cos 2\theta
\end{eqnarray}
where $\theta = \arctan(({y_{\rm source}-y_{\rm lens}})/({x_{\rm source}-x_{\rm lens}}))$.  (x$_{\rm lens}$, y$_{\rm lens}$) and (x$_{\rm source}$, y$_{\rm source}$) are the coordinates of the lens and source, respectively.

The average 45$^{\circ}$-rotated ellipticity component is calculated as a systematic test as it is expected to be consistent with zero at all radial scales. Two additional checks for systematics include measuring the lensing signal of stars and the lensing signal of source galaxies around random locations in the catalog. The results of these systematics checks, all consistent with a null result, are shown along with the mean galaxy-galaxy lensing signal for all galaxy lenses in the redshift range 0.35 $<$ $z$ $<$ 0.55 and absolute magnitude range -22 $<$ M$_{R}$ $<$ -19 in Figure~\ref{F:meansys}.

\subsection{\label{photozsec}Photometric Redshifts}

\begin{figure}
\includegraphics[width=\hsize]{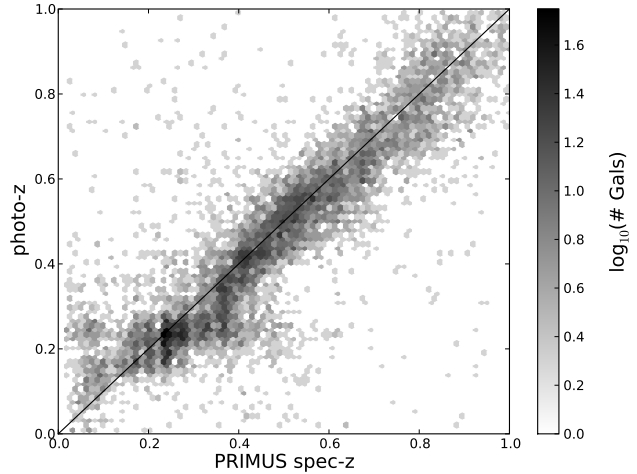}
\caption{\label{F:zspecvszphot} Density plot of spectroscopic z vs photometric z.  The photo-z are scattered about the black solid line that represents the spec-z=photo-z relation with $\sigma_{photo-z}=0.06(1+z_{spec-z})$.  Regions where there is notable bias include galaxies at spec-z$\sim$0.1 that are scattered to larger photo-z and galaxies at 0.3$<$spec-z$<$0.5 that are assigned to photo-z$\sim$0.25.  This plot shows that we must be careful with photo-z$<$0.3 and take relatively wide vertical photo-z slices in our analysis of at least 0.2.}
\end{figure}

We compute photometric redshifts (photo-z) by fitting the four-band photometry to a set of galaxy templates to determine both redshift and galaxy type. This is done with the publicly available Bayesian Photometric Redshift code \citep[BPZ;][]{2000ApJ...536..571B}. The code utilizes a Bayesian prior for the probability of the redshift given the type and magnitude.  Instead of using the default BPZ magnitude prior which treats galaxies with R$<$20 identically and is based on a small number of galaxies in the Hubble Deep Field North \citep[HDFN;][]{1996AJ....112.1335W}, we fit new parameterized priors following the prescription in \citet{2000ApJ...536..571B}. For R$<$21, we calibrate the prior with spectroscopically confirmed galaxies in the Smithsonian Hectospec Lensing Survey \citep[SHELS;][]{2005ApJ...635L.125G}, which is complete to R=20.7.  For fainter galaxies, we fit to the VIMOS-VLT Deep Survey \citep[VVDS;][]{2005A&A...439..845L}, which consists of more than 11,000 spectra from 17.5$<$i$_{\rm AB}<$24.0.  While the possibility remains that there is a population of galaxies systematically missing from our magnitude complete training sets that could introduce a bias in the calculated prior, we only broadly fit for three galaxy classes (Elliptical, Spiral, and Starburst) and use a parameterized N(z), which mitigates small missing populations. The resulting prior is qualitatively very similar to the HDFN prior but is based on almost 100 times as many spectroscopic redshifts and has a more sophisticated description for R$<$20.  We empirically adjust the six standard CWW+SB \citep{1980ApJS...43..393C,1996ApJ...467...38K} templates using the photometry of galaxies with known spectroscopic redshifts (spec-z) in SHELS, and we employ the resulting templates to determine photo-z and K-corrections to $z=0$.  In Figure~\ref{F:zspecvszphot}, we show a number density plot of spec-z vs photo-z for 10,000 galaxies in the southern field F5, which has overlap with the PRIsm MUlti-object Survey \citep[PRIMUS; ][]{2011ApJ...741....8C}.  The quoted 100\% sampling range is R$<$22.8, the 30\% sampling range is 22.8$<$R$<$23.3, and the redshift precision is $\sigma_{spec-z}=0.005(1+z_{true-z})$ due to the low spectral resolution of the instrument.  
The root-mean-square $\sigma_{\rm z,rms}$ of the difference between the observed redshift and the true redshift of a galaxy at each redshift bin is defined as:
\begin{equation}
\sigma_{\rm z,rms}^2 = {\left\langle \Bigg( \frac{z_{\rm obs}-z_{\rm true}}{1+z_{\rm true}}\Bigg)^{2} \right\rangle} 
\end{equation}
After outlier rejection, the DLS photo-z scatter $\sigma_{photo-z}=0.06(1+z_{spec-z})$, and about 4\% of the galaxies have photo-z outside of $0.2(1+z)$ (catastrophic outliers).  Further details can be found in Schmidt \& Thorman (2012, in prep.).

BPZ reports both a redshift probability density function (PDF) and a one point summary statistic z$_{B}$, the peak of the PDF. The PDF is the posterior distribution given the data, in a Bayesian sense, marginalized over template type and apparent magnitude. In this work, we use z$_{B}$ when calculating Equation~\ref{eq:sigcriteq}.  We also use z$_{B}$ in order to calculate absolute magnitudes and rest-frame luminosities for the lens galaxies, differentiate the lens sample from the source sample, and weight the galaxies by the lensing kernel. We use the terms z$_{B}$ and photo-z interchangeably for the remainder of the paper.

\begin{figure}
\includegraphics[width=\hsize]{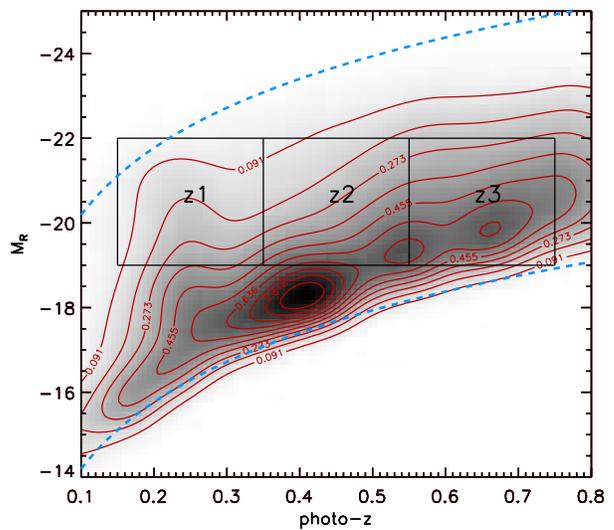}
\caption{\label{F:zvsM} Photo-z vs absolute R magnitude for redshift and magnitude selected lenses. The black contours indicate the number density, with levels given by logarithmically spaced bins from 1 to 14,000.  The dashed blue lines correspond to the faint apparent magnitude cutoff ${\rm m}_{R,{\rm max}}=24$ and the bright apparent magnitude cutoff ${\rm m}_{R, {\rm min}}=18$. The black rectangles show the cuts we use in Section~\ref{zevolsect} when we make comparisons between redshift shells, labeled z1-3.}
\end{figure}

Table~\ref{tab:lens_src_cuts} lists the redshift and magnitude selections applied to create lens and source samples. All source galaxies have $m_{R}<24.5$. This conservative faint cut-off corresponds to the peak of the apparent magnitude number counts for the joint catalog containing galaxies with both shapes and photo-zs.  After further conservative cuts in photo-z and magnitude described in Table~\ref{tab:lens_src_cuts} as well as data quality and S/N cuts, the trimmed source catalogs used in this analysis contains between 225,580 and 506,241 galaxies.  The corresponding luminosity-binned lens sample contains 436,943 galaxies. We plot number density contours for photo-z vs absolute R magnitude in Figure~\ref{F:zvsM} for the redshift range 0.1$<$z$_{B}$$<$0.8 and 18$<$m$_{R}<$24. The three black squares indicate the volume-complete lens samples that we later compare in Section~\ref{zevolsect}.

\begin{figure}
\includegraphics[width=\hsize]{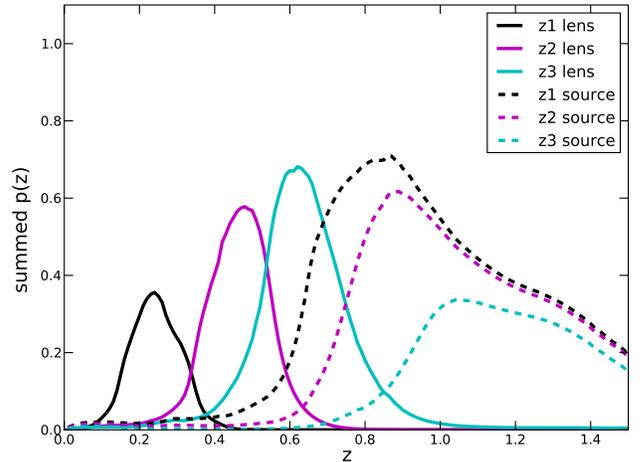}
\caption{\label{F:z1z2pofz} The summed normalized likelihood function p(z) from BPZ for redshift and magnitude selected sets of lenses and sources. The cuts applied to create the lens and source samples are described in Table~\ref{tab:lens_src_cuts}.
The lens and source samples are generally well-separated. However, there are non-negligible overlaps between the z2 lenses and sources and between the z1 lenses and z2 lenses. We address the effect of these overlaps on our analysis using mock catalogs, and angular cross-correlations within the observational data,  and discuss our findings in Section~\ref{photozsim}.
} 
\end{figure}

To check our lens and source selections using z$_{B}$, we show in Figure~\ref{F:z1z2pofz} the arbitrarily normalized summed $p(z)$ from BPZ for foreground lenses selected in a fixed absolute magnitude range 
-19$<$M$_{R}$$<$-22 and split into redshift shells of 0.15$<$z$_{B}$$<$0.35, 0.35$<$z$_{B}$$<$0.55, and 0.55$<$z$_{B}$$<$0.75 and for background source samples selected such that their peak p(z) occurs at more than 0.3 higher redshift than the peak in the source sample (see Table~\ref{tab:lens_src_cuts} for the specific cuts).  
%0.6$<$z$<$1.5 and 0.9$<$z$<$1.5 
The resulting summed $p(z)$ show lens distributions that are generally well-separated from the source distribution.  We consider the effects of photo-z inaccuracy including the residual systematic errors due to the $\sim$4\% outliers on the lensing analysis in Section~\ref{photozsim}.

\begin{table}[t]
\centering
\begin{tabular}{lccccccc}
\hline
Ref &z$_{-}$ &z$_{+}$ &M$_{R,-}$ &M$_{R,+}$ &m$_{R,-}$ &m$_{R,+}$ &N$_{gal}$\\
\hline
L1 & 0.35 & 0.75 & -24 & -22 &18 &24 &38,505\\
L2 & 0.35 & 0.75 & -22 & -20 &18 &24 &178,227\\
L3 & 0.35 & 0.75 & -20 & -18 &18 &24 &220,211\\
SL & 0.75 & 1.5 & n     & n     &22&24.5 &225,580\\
z1 & 0.15 & 0.35 & -22 & -19 &18 &24 &61,527\\
z2 & 0.35 & 0.55 & -22 & -19 & 18 &24 &124,744\\
z3 & 0.55 & 0.75 & -22 & -19 & 18 & 24 &188,908\\
Sz1 &0.65 & 1.5 &n &n &22 &24.5 &506,241\\
Sz2 & 0.75 & 1.5 &n &n &22 &24.5 &402,339\\
Sz3 &0.95 & 1.5 &n &n &22 &24.5 &225,580\\
\hline
\end{tabular}
\caption{Summary of the cuts applied to select the lenses L1-L3, z1-z3 and the corresponding sources SL and Sz1-Sz3.  "-" means minimum and "+" means maximum.  "n": no threshold imposed.}
\label{tab:lens_src_cuts}
\end{table}

\subsection{Star/Galaxy Separation}

For star/galaxy separation, we apply two cuts.  At bright magnitudes, stars are easily distinguishable from galaxies in a size vs magnitude diagram. We use SExtractor FLUX\_RADIUS as our measure of size, and cut out objects with FLUX\_RADIUS$<$2.4 pixels with m$_{R}<$19.  The pixel size is 0.257 arcseconds.  At fainter magnitudes, the star and galaxy loci overlap on the size vs magnitude diagram, so we employ a cut which involves all three second central moments of the PSF and the object.  For each object, we determine the total $\chi^2$ for the hypothesis that it is a point source, using the three second moments of the object and of the PSF at that location.  We then cut on $\chi_{tot}^2>5$ to eliminate most faint stars.  This procedure was shown to result in $\sim 0.4\%$ stellar contamination in spectroscopic targets for m$_{R}<22.5$ \citep{2011arXiv1110.4391D}.  At fainter magnitudes, the number density of galaxies overwhelms that of stars while the efficiency of the $\chi^2$ cut should not diminish (although a larger fraction of galaxies might be cut).  Using stellar population models from \citet{2003A&A...409..523R} and DLS coordinates to estimate star counts and typical galaxy counts from the literature,\footnote[2]{See http://star-www.dur.ac.uk/$\sim$nm/pubhtml/counts/counts.html} we estimate the stellar contamination to be $<0.2\%$ for  m$_{R}\sim 24$.  The shape cuts described in Section~\ref{shapes} further reduce this stellar contamination.

\section{Photometric Redshift Error Analysis}

Photo-z errors can affect the amplitude of the galaxy-mass correlation in several ways, which we can separate into those related to errors in lens galaxy photo-z ($z_{L}$) and those related to errors in source galaxy photo-z ($z_{S}$).  First consider errors in lens photo-z.  Error in $z_{L}$ will smear the signal in projected comoving radius due to error in the conversion from an angular distance to a comoving distance.  Especially for sources close to the lens redshift there will also be an error in $\Delta \Sigma$ due to an error in $\Sigma_{\rm crit}$ (see Equation~\ref{eq:dsig_calc}).  In the case that $z_{L} > z_{S}$, the signal will be diluted by ``source'' galaxies which are in reality foreground objects.  There is an additional issue that the errors in $z_{L} $ propagate to errors in absolute magnitude M$_{R}$.  As found by e.g. \citet{2005ApJ...635...73H}, the net effect (due to the shape of the galaxy number counts) is that intrinsically fainter galaxies are more often upscattered into brighter luminosity bins, causing a dilution in the observed signal.  Now consider errors in source photo-z.  For a source galaxy that is behind a lens galaxy an error on a source redshift will create a bias in $\Delta \Sigma$ due to an error on $\Sigma_{\rm crit}$.  An error in a source redshift such that $z_{S} < z_{L}$ will create a signal dilution because this "source" galaxy is now a foreground object.  The net correction is a mix of all these effects, and the mix changes with galaxy and lens samples.  For further discussion see Figure 8 of \citet{2010ApJ...709...97L} and Section 7 of \citet{2012MNRAS.420.3240N}.  

We undertake a galaxy-galaxy lensing Monte Carlo simulation, creating mock lens and source catalogs representing DLS-sized fields of 4 deg$^{2}$ to understand how photo-z scatter, photo-z catastrophic outliers, and lens/source sample selection affect the accuracy of the projected galaxy-mass correlation measurement.  We also compute angular cross-correlation between the lens and source samples to investigate the purity of these samples.

\subsection{\label{photozsim} Signal Recovery Simulations}

We calculate the ratio of observed (with scattered redshifts) lensing signal to input lensing signal $\Delta \Sigma (R)_{\rm scatt} / \Delta \Sigma (R)_{\rm input}$ for each of the three lens sample redshift cases.  For the mock source galaxies we draw random x and y positions for simulated source galaxies, and we draw redshifts and intrinsic ellipticities in a Monte Carlo fashion from distributions consistent with the data.  We draw the intrinsic ellipticity components along the major and minor axes from Gaussians with width 0.3 and centered at 0. The summed observed p(z) for sources (as shown in Figure~\ref{F:z1z2pofz}) is a broadened approximation for the true dN/dz of the sources since this distribution already includes observational effects such as photo-z errors and is broader than the underlying distribution. To simulate the native distribution of galaxies in redshift for each galaxy population, we find that distribution p0(z) which when convolved with the known $\sigma_z$ photo-z errors gives the observed summed p(z) for that population.  We draw redshifts from the summed intrinsic p0(z) distribution of the sources. 

For the mock lens galaxies, we fix the x, y, and luminosity of the simulated lens galaxies to the values measured from the observed catalogs.  That is, each simulated lens galaxy has a one-to-one counterpart in the real catalog (e.g. that of F5) from which it inherits x and y coordinates in pixels and luminosity.  These mock lens galaxies thus inherit the angular clustering that is present in the observed sample of lens galaxies. As with the sources, we draw redshifts for the lenses from an estimate of the intrinsic summed p0(z) of the lenses which when convolved with the photo-z error yields the observed p(z) as shown in Figure~\ref{F:z1z2pofz}. We also take into account any redshift overlap between the lenses and sources and among the z1, z2, and z3 lens bins. To convert the lens luminosities to halo masses consistent with the data, we fit NFW profiles to the inner 300 kpc of the lensing signal in each luminosity bin and fit a power law to the mean luminosity vs halo mass as shown below in Figure~\ref{F:ML_PL}. The best fit mass-luminosity relation is:
\begin{eqnarray}\label{eq:dls_ML}
  M_{200} = 1.46 \times 10^{3} M_{\odot} ~ \langle L / L_{\odot}\rangle^{0.89}
\end{eqnarray}
To mitigate edge effects, we pad the boundaries of each simulated mass field with an additional degree on each side. That is, the lens galaxies are distributed over a total area of 16 deg$^{2}$ for each of the five fields.

Using NFW mass profiles for the lens halos, we calculate the expected shear for each lens-source pair given the physical separation and lens $M_{200}$.
For each source galaxy $i$, we use the weak lens approximation in calculating the total applied shear due to all lens galaxies $\gamma_{T,i}=\sum_{j}^{N_{\rm lens}}\gamma_{j}$.  The observed ellipticity is then given as in Equation~\ref{eq:obsell}. The resulting mock lens and source catalogs have the selection effects of our observations, and is a starting point from which to test how photo-z errors such as scatter, photo-z bias, and catastrophic outliers affect the galaxy-mass correlations reconstructed from the lensing analysis via the corresponding errors in the distance ratios. To examine this, we assign ``observed redshifts'' by applying photo-z errors similar to those measured from the observational data.   We obtain the observed redshifts by drawing from a Gaussian centered at the assigned true redshift with width given by a value describing the photo-z scatter (see Section~\ref{photozsec}).  We use a photo-z scatter of  $\sigma_{\rm z,rms}$=0.06(1+z).  We also introduce catastrophic outliers by switching the redshifts of a random 4\% of the galaxies with random redshifts.

After determining the absolute magnitudes corresponding to the scattered redshifts, we re-select lenses and sources using these new ``observed'' redshifts and absolute magnitudes and measure $\Delta \Sigma (R)_{\rm scatter}$ as in Equation~\ref{eq:dsig_calc}.
At very large angular separation there is no mechanism that can cause a spatial dependence of the ratio $\Delta \Sigma (R)_{\rm scatt} / \Delta \Sigma (R)_{\rm input}$, and we fit a constant over 1-10 Mpc radius.  We repeat this entire simulation for various levels of photo-z error and find that the recovery ratio for each lens redshift sample decreases with photo-z error. 
The best fits for this recovery ratio for 0.06(1+z) scatter are 0.81, 0.85, and 0.90 for L1, L2, and L3, respectively and 0.94, 0.83, and 0.74 for z1, z2, and z3, respectively.  In the analysis in Section 7.3 we correct for this systematic error.

\subsection{Angular Cross-Correlations}

Do the summed p($z$) distributions for our lens and source samples represent the true distributions?  While the distributions of PRIMUS spec-z, DLS photo-z, and p(z) are consistent out to the limit of the spectroscopy at $z \sim 1$, it is important to investigate consistency for our entire source and lens sample galaxies in other ways.  As a test of source and lens sample purity we calculate angular cross-correlations ${\rm w}_{12}(\theta)$ between these samples.  We find that the ratios between the auto and cross correlations of these samples is less than 0.1, consistent with the overlap of the p($z_{\rm 1,2}$) tails and the known photo-z outlier rates.  The amplitude of ${\rm w}_{12}(\theta)$ never rises above 0.01.  The residual sample impurity results in a slight decrease in signal-to-noise ratio, but is small enough that we do not make any additional corrections in this analysis.

\section{Results}

We measure the cross-correlation between the lens galaxies and the associated total projected mass distribution which is in the foreground of the source galaxies. The DM halo in which the galaxy is resident, as well as all other projected mass correlated with the galaxy, contribute to this weak lens signal.
We bin the shear signal in logarithmic bins by projected comoving radii and convert to a projected differential surface mass density following Equation~\ref{eq:dsig_calc}.  Errors are calculated via jackknife resampling where the jackknife components are one ninth of a field size. Over the 5 fields, this yields a total of 45 jackknife subsamples.  These data permit an analysis of the distribution of mass associated with foreground galaxies over a wide range of projected physical scales, and its correlation with galaxy luminosity and redshift.

\subsection{\label{lumsect}Galaxy-Mass Correlation vs Luminosity}

\begin{figure}
\includegraphics[width=\hsize]{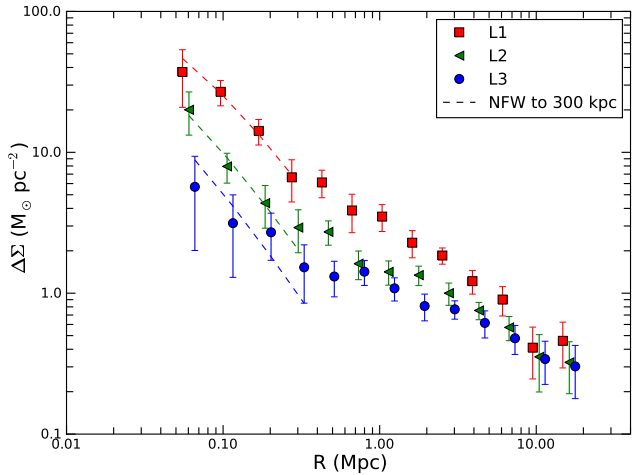}
\caption{\label{F:lum} Galaxy-mass correlation (expressed as a mass overdensity around lens galaxies) vs radius and luminosity.  The surface mass overdensity for three bins of lens absolute luminosity as a function of radial separation is shown.  All galaxies have 0.35$<$z$_{B}$$<$0.75.  Red squares show L1, green triangles show L2, blue circles show L3, and the NFW halo model fits to the inner 300 kpc are overplotted as dashed lines.  Correction factors from photo-z errors, as discussed in Section~\ref{photozsim}, have been applied.  L2 and L3 data points are slightly horizontally offset for clarity.  On small radial scales, the profiles scale with luminosity such that the brighter lensing galaxies have higher galaxy-mass correlation. On large radial scales, the profiles flatten, indicating additional mass contributions from neighboring halos. The shape of these radial profiles and trends with luminosity are similar to the theoretical prediction shown in Figure 8 of \citet{2008MNRAS.388....2H}.}
\end{figure}

We divide the lenses into three ranges of absolute magnitude for a fixed range of 0.35$<$z$_{B}$$<$0.75 and measure the galaxy-galaxy lensing signal for each lens galaxy luminosity sample.  The properties of these samples are summarized in Table~\ref{tab:lens_src_cuts}.  After dividing the signal by the correction factors due to photo-z errors discussed in Section~\ref{photozsim} (0.81, 0.85, and 0.90 for L1, L2, and L3, respectively), the results are shown in Figure~\ref{F:lum} along with NFW profiles fit to the inner 300 kpc of each $\Delta \Sigma(R)$.  These fits take into account the full covariance matrices estimated from the data (see the Appendix for details about how the covariance matrix is calculated), although this inclusion does not have a significant effect due to our focus on small radial scales.  The signals scale with average luminosity with higher surface mass overdensities corresponding to brighter average luminosities.   As expected, intrinsically more luminous galaxies have higher mass profiles inside their virial radii than low luminosity galaxies.

As shown in Figure~\ref{F:lum}, on scales larger than $\sim 0.3$ Mpc all three signals flatten, and converge on 10 Mpc scales to a narrow range of mass overdensity.  This is because at large radii the galaxy-mass correlation reflects the mass auto-correlation modulo the galaxy bias, as discussed in \citet{2008MNRAS.388....2H}.  \citet{2008MNRAS.388....2H} use the Millenium simulation to examine the cross correlation between halo centers and mass, showing a dependence on galaxy mass (or luminosity) similar in profile shape to our results (their Figure 8).  The strong luminosity dependence of $\Delta \Sigma$ at R$<$1 Mpc and weak luminosity dependence at R$>$1Mpc have also been found in the simulations of \citet{2004ApJ...614..533T} and the observations of \citet{2004AJ....127.2544S}.  The behavior of the galaxy-mass correlation at large radii has also been recently examined by \citet{2012ApJ...746...38M} in a $1024^3$ particle N-body CDM simulation.  

\subsection{\label{massvslum}Mass vs Luminosity}

From the best fit NFW profiles, we obtain values for M$_{200}$ for each luminosity bin.  It is worth noting that the NFW is not a good fit to the $\Delta \Sigma$ profiles beyond a few hundred kpc.  In particular, the L3 sample likely contains a non-negligible fraction of satellite galaxies, and preliminary work fitting a full halo model indicates M$_{200}$ for this luminosity bin is overestimated by $\sim$20\%.

As pointed out in \citet{2004ApJ...614..533T}, the best fit NFW mass generally falls between the mean and median mass for a broad halo mass distribution.  The conversion from the best fit NFW mass to the mean mass is an upwards correction that depends on the scatter in the mass-luminosity relation, which is higher at the bright end where the halo mass distribution is broader.  To determine this correction, we follow the procedure described in \citet{2011arXiv1107.4093V}.  We adopt a conditional probability function for the halo mass given a luminosity of the form:
\begin{equation}
P(m_{\rm halo} | l) \propto \exp \left( -\frac{(m_{\rm halo} - m_{\rm halo, cent})^{2}}{2\sigma^{2}_{m_{\rm halo}}} \right)
\end{equation}
where $l = \log(L)$, $m_{\rm halo} = \log(M_{\rm halo})$, and $\sigma^{2}_{m_{\rm halo}}$ is the scatter in $m_{\rm halo}$.  For a given best fit NFW mass and a $\sigma_{m_{\rm halo}}$ from work on satellite kinematics in SDSS by \citet{2011ApJ...741...19M} (their Figure 4 produces values from 0.35 to 0.45 for our mean luminosities), we then convolve the conditional probability function with the mass function from \citet{2008ApJ...688..709T} and draw 1000 masses from the resulting distribution.  We calculate and average the NFW $\Delta \Sigma (R)$ profiles for this ensemble of masses and compare the best fit NFW M$_{200}$ to $\langle M_{200}\rangle$.  We obtain conversion factors of 3.8 (L1), 3.1 (L2), and 2.5 (L3) and multiply our best fit NFW M$_{200}$ by these numbers to obtain $\langle M_{200}\rangle$.

\begin{figure}
\includegraphics[width=\hsize]{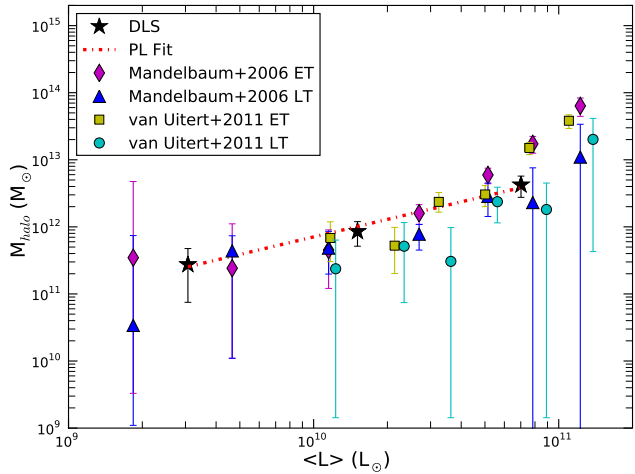}
\caption{\label{F:ML_PL} The mean luminosity vs mean halo mass for the three luminosity bins described in Table~\ref{tab:lens_src_cuts}.  Halo masses calculated by fitting the NFW profile to the inner 300 kpc are shown as black stars and then using the conversion factors described in Section~\ref{massvslum} to obtain mean masses.  The dashed red line shows a power law fit to the mean masses. The power law given by Equation~\ref{eq:dls_ML} is used in the Monte Carlo simulations discussed in Section~\ref{photozsim} to convert lens luminosities to halo masses. For comparison purposes, previous results are also plotted, although note that different methodologies are used. \citet{2006MNRAS.368..715M} results at $\langle z\rangle\sim$0.1 are shown as blue triangles for late-types (LT) and thin magenta diamonds for early-types (ET).  Note that we have divided the data points from \citet{2006MNRAS.368..715M} by 1.3 to correct for different mass definitions. \citet{2011arXiv1107.4093V} results at $\langle z\rangle\sim$0.1 are shown as cyan filled circles for late-types and yellow squares for early-types.  The errors on the SDSS results are 2-$\sigma$. The DLS results agree well with those in the literature after making adjustments for mass definitions and passive luminosity evolution as discussed in the text.}
\end{figure}

We plot $\langle M_{200} \rangle$ vs the mean luminosity of each bin in Figure~\ref{F:ML_PL} along with a simple power law fit (Equation~\ref{eq:dls_ML}).  In Figure~\ref{F:ML_PL}, our mean luminosities are converted to the AB system.  Since the mean redshifts of the three luminosity bins decrease slightly with decreasing luminosity, we also scale the DLS masses to the mean redshift of L1 (z$=$0.59) for the sake of consistency.  The mean redshifts of L2 and L3 are 0.58 and 0.51, which corresponds to multiplication factors of 1.01 and 1.10 based on our M$_{200}$ definition which depends on redshift through $\rho_{\rm crit}$.  We perform the power law fit primarily for use in the Monte Carlo simulations described in Section~\ref{photozsim}.  We also show results for early-type (ET) and late-type (LT) galaxies at low redshift from \citet{2006MNRAS.368..715M} and \citet{2011arXiv1107.4093V}.  However, there are several caveats to consider before making a comparison.  First, \citet{2006MNRAS.368..715M}  use a different mass definition (180$\bar{\rho}$ instead of our 200$\rho_{c}$), so our masses should be adjusted upwards by $\sim$30\% for a more direct comparison.   In Figure~\ref{F:ML_PL}, we plot the \citet{2006MNRAS.368..715M} masses after dividing by 1.3 to make the qualitative comparison.  Second, the DLS points are at a redshift of 0.59, and we have not yet included corrections for passive evolution.  If we adopt a passive evolution correction from \citet{2003AJ....125.2348B} of $1.6(z-0.1)$ for galaxies that are best fit with elliptical templates (e.g. BPZ T\_B$<$1.5),  K-correct to $z=0.1$, and calculate the distance modulus with $h=1.0$ as in \citet{2005MNRAS.361.1287M}, our mean luminosities decrease by $\sim$30\%.  Third, we have not split our lens samples by type, so the DLS masses include both ET and LT galaxies.  Finally, the conversion factors we applied to obtain the conversions from best fit NFW M$_{200}$ to $\langle M_{200}\rangle$ are based on our adopted $\sigma_{m_{\rm halo}}$ from \citet{2011ApJ...741...19M} whose results are limited to z$\sim$0 central galaxies.  The true $\sigma_{m_{\rm halo}}$ for our higher redshift central and satellite galaxies might be somewhat different, ultimately changing our interpretation of $\langle M_{200} \rangle$.

\subsection{\label{zevolsect}Galaxy-Mass Correlation vs Redshift}

\begin{figure}
\includegraphics[width=\hsize]{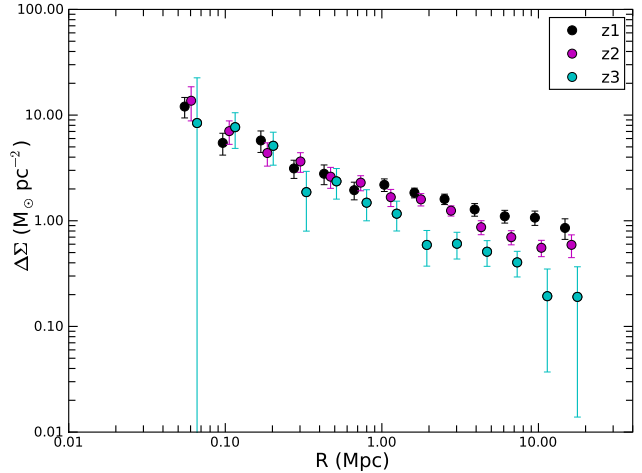}
\caption{\label{F:gglz1z2L123} Redshift dependence on all scales for a single wide range of luminosities -22$<$M$_{R}<$-19. We compare the galaxy-mass correlation vs radius for three lens redshift shells, which have properties described in Table~\ref{tab:lens_src_cuts}.  These signals include corrections for photo-z errors that were derived from mock catalogs as discussed in Section~\ref{photozsim}.  z1 is shown in black, z2 in magenta, and z3 in cyan.  The z2 and z3 points are slightly horizontally offset for clarity.  On small radial scales, the three samples reflect signals consistent with a similar mass.  However, on larger scales, there is a trend where higher $\Delta \Sigma$ corresponds to lower mean redshift.}
\end{figure}

We next divide the lenses into three ranges of redshift for a fixed range of absolute magnitude -22$<$M$_{R}<$-19 and measure the galaxy-galaxy lensing signal for each redshift sample.  This range of absolute magnitude was chosen such that the samples could be as volume complete as possible (see Figure~\ref{F:zvsM}) while still allowing a reasonable S/N.  The properties of these samples are summarized in Table~\ref{tab:lens_src_cuts}, and the results are shown in Figure~\ref{F:gglz1z2L123} after applying the corrections for photo-z errors found in Section~\ref{photozsim}: 0.94 (z1), 0.83 (z2), and 0.74 (z3).  The bins at large separations are correlated, and we discuss how we calculate the covariance matrices in the Appendix and show the normalized covariance matrices in Figure~\ref{F:corrmatrix}.  At radii less than a few hundred kpc, all three signals have similar amplitudes, indicating that the mass range corresponding to the luminosity selection is consistent across redshift shells.  However, there is a noticeable trend at larger radii where the amplitude of $\Delta \Sigma$ increases with decreasing redshift.  Both before and after applying the recovery corrections from Section~\ref{photozsim}, there is a trend.

\section{Discussion}

We have presented projected galaxy-mass correlations measured using galaxy-galaxy lensing over a wide baseline in radius, luminosity, and redshift.  On small radial scales, these signals probe the individual DM halos in which galaxies reside.  On larger radial scales, these signals are sensitive to correlated large-scale structure and neighboring DM halos.  We have focused on two lines of investigation:  fixing the redshift range and varying the mean rest-frame luminosity and conversely, fixing the luminosity range to a volume-complete sample and varying the mean redshift.

For fixed redshift and varied luminosity, we find the well-established scaling of the halo mass with luminosity: intrinsically brighter galaxies are also more massive.  In Figure~\ref{F:ML_PL}, we show a comparison of our results for the M$_{halo}$-L relation at z$\sim$0.6 with those of \citet{2006MNRAS.368..715M} and \citet{2011arXiv1107.4093V} at z$\sim$0.1.  After corrections for differing mass definitions and passive luminosity evolution, we find reasonable agreement with the literature results.

On larger radial scales, the shape of the galaxy-mass correlations and their dependence on galaxy luminosity look quite similar to the simulation results of \citet{2008MNRAS.388....2H}. There are currently few lensing observations of galaxies over wide ranges of luminosity and redshift at radii greater than a few Mpc. We can qualitatively compare our results shown in Figure~\ref{F:lum} with four observations at lower redshift.  Using shear measurements around z$\sim$0.1 SDSS galaxies, \citet{2004AJ....127.2544S} found trends with luminosity that are similar to the higher redshift findings presented here; however, they fit power laws to their shear and galaxy-mass correlations and did not see significant deviation from power laws at large scales.  We can also compare qualitatively to \citet{2010Natur.464..256R}, who measure shear around luminous red galaxies (LRG) in SDSS out to large radii and model it with a halo model. Their LRG mass overdensity at z=0.3 is consistent with our low redshift z1 and z2 samples at 10 Mpc.  \citet{2011arXiv1107.4093V} study galaxies in RCS2 with spectroscopic redshifts from SDSS, limiting the lens redshifts to z$\sim$0.1.  Their Figure 8 shows halo model fits to the lensing signal out to 10 Mpc, which is consistent with the signal we measure.  In these previous studies, the lens galaxies are closest in mass to our L1 sample, but are generally more massive.  Finally, \citet{2012arXiv1207.1120M} show results for $\Delta \Sigma$ for three redshift bins out to z$\sim$0.5.  However, a direct comparison is difficult because they probe much more massive and more highly biased galaxies (i.e. LRGs).

Most interestingly, our galaxy-galaxy lensing data allows an investigation of the redshift evolution of the projected galaxy-mass correlation on 10 Mpc scales at fixed lens galaxy luminosity. We find evidence for growth over time of the galaxy-mass correlations on large scales from z$=$0.65 to z$=$0.2 as shown in Figure~\ref{F:gglz1z2L123}.
Interpretation of this result in terms of evolution of bias and growth of LSS mass structure with cosmic time requires additional information. The overall effect as a function of lens sample redshift depends on both LSS growth and galaxy bias as a function of mass and type, and an additional probe such as the lens galaxy clustering signal is necessary to disentangle the growth and bias.  We will combine this two-point auto and cross-correlation information for each of the three lens samples with the galaxy-mass correlations reported here in a joint analysis in future work.

\section{Acknowledgments}

We thank Ian Dell'Antonio, Paul Thorman, and Russell Ryan for help with data reduction, Lloyd Knox, Jeff Newman, Michael Schneider, and Catherine Heymans for useful discussions, Andrew Bradshaw for efforts on the shape measurement, and Perry Gee for technical support.  We thank the anonymous reviewer for their helpful suggestions. This research was funded in part by  DOE grant DE-FG02-07ER41505, NSF AST-1108893, UC Davis, and the TABASGO Foundation.  AC acknowledges support from the European Research Council under the EC FP7 grant number 240185.  CM acknowledges support from NSF grant AST-1009514.
Funding for the Deep Lens Survey has been provided by Lucent Technologies and NSF grants AST 04-41072 and AST 01-34753. Observations were obtained at Cerro Tololo Inter-American Observatory and Kitt Peak National Observatory.  CTIO and KPNO are divisions of the National Optical Astronomy Observatory (NOAO), which is operated by the Association of Universities for Research in Astronomy, Inc., under cooperative agreement with the National Science Foundation.

\bibliography{ms_bib}{}
\bibliographystyle{apj}

\newpage

\appendix

\section{Systematics Tests}
\label{sec:systest}

\begin{figure}
\centering
\includegraphics[width=4.0 in]{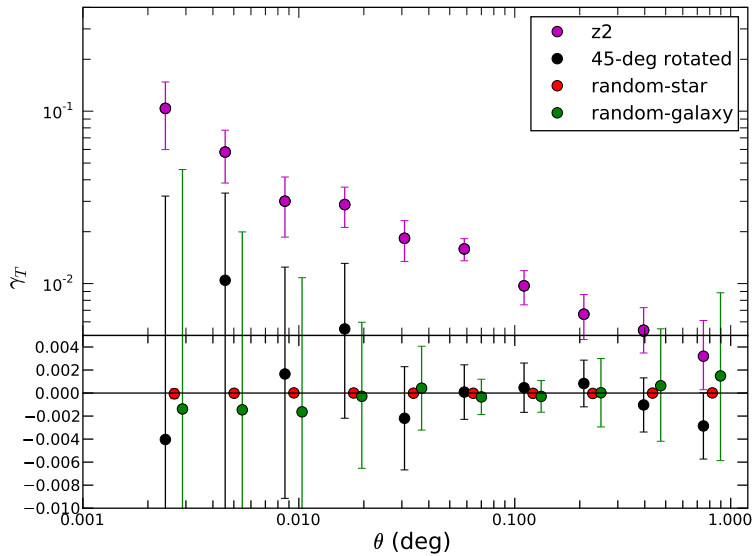}
\caption{\label{F:meansys} The mean tangential shear for the z2 lenses and three systematics checks as a function of angular scale. Note that the y-axis is log-scale down to 0.005 and linear-scale below in order to magnify the small systematics test results.  The black circles represent the signal from 45$^{\circ}$-rotated source shears, the red circles show the signal from one million random positions cross-correlated with star shears, and the green circles show the signal from one million random positions cross-correlated with source galaxy shears.  The data points are offset horizontally for clarity.  Errors are estimated from jackknife resampling, and the errors on the random-star cross-correlations are too small to be visible on the plot.  The systematics tests are all consistent with zero, and generally at least an order of magnitude below the mean tangential shear of the real foreground galaxy positions cross-correlated with background galaxy shears (magenta circles).}
\end{figure}

We undertake several tests for shear and photo-z systematics. Figure~\ref{F:meansys} shows the results of three checks for shear systematics as a function of angular scale: foreground lens positions cross-correlated with tangential shear from 45-degree rotated source galaxies, one million random foreground positions (200k per field) cross-correlated with the tangential shear from source galaxies, and one million random foreground positions cross-correlated with the tangential shear from stars. In order to plot the positive and large shear signal on the same plot with the relatively small residual bipolar systematics, the y-axis is log above 0.005 shear and linear below.  These tests are consistent with zero residual shear systematics and are typically less than the true signal by at least an order of magnitude.  The error bars are given by jackknifing 9 subsamples for each of the 5 fields.  The two tests using actual source galaxy shears have the largest errors since there is a much larger variation in the galaxy ellipticity distributions than for the stars, which should have minimal variation by construction.  Note that on large scales, the errors become correlated.

\begin{figure}
\centering
\includegraphics[width=4.0 in]{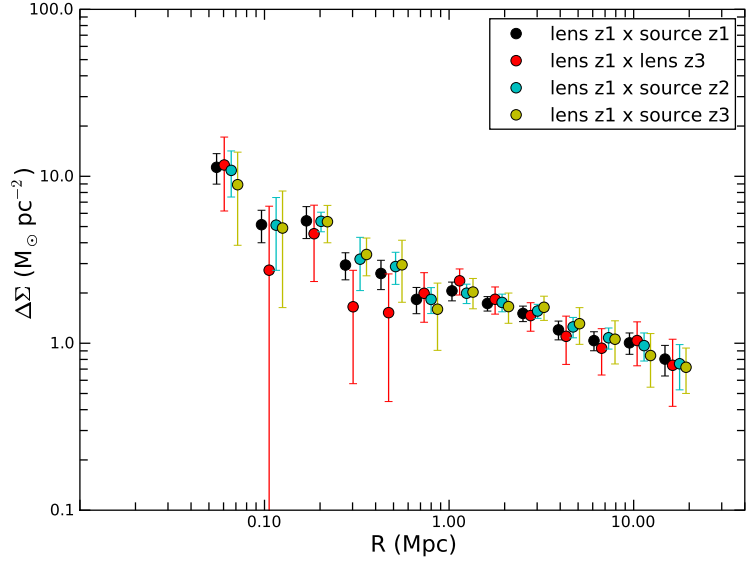}
\caption{\label{F:scalingsys} Lensing of z1 by source samples with different mean redshift.  We show a test of photo-z and lensing consistency by measuring the galaxy-mass correlation in one lens galaxy sample by using all the possible source samples.  $\Delta \Sigma$(R) for the z1 lens sample is cross-correlated against four source shells with very different mean redshifts, described in the legend. 
The black circles show the cross-correlation with the z1 background sources as shown in Figure 7, the red circles show the cross-correlation with z3 lenses (used in this test as a background sample), the cyan circles show cross-correlation with the z2 background sources, and the yellow circles show the cross-correlation with the z3 background sources.  
Table 1 describes the cuts for each of these samples.  The data points are calculated for the same radial bins but are offset for clarity.  While the four signals are correlated with each other, the mean redshift of the background sample varies.  The distance ratio scaling appears to be consistent.}
\end{figure}

If either the photo-z of the lens or source galaxy samples are systematically in error, then the error in the distance ratio propagates to an error in the galaxy-mass correlation.  Such an error would be revealed by differences in amplitude in the galaxy-mass correlations computed using different source samples for the same lens sample.  Figure~\ref{F:scalingsys} shows the cross-correlation of the z1 lens sample with four source samples with differing mean redshifts.  We find no evidence of a distance ratio scaling inconsistency at the level of the noise.

\section{Covariance Matrices}

\begin{figure}
\centering
\includegraphics[width=7.0 in]{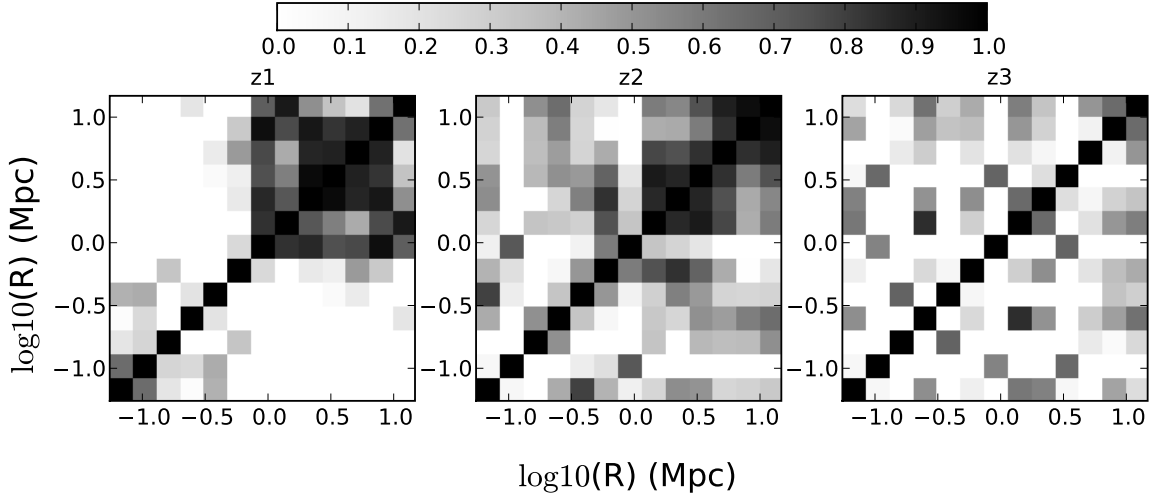}
\caption{\label{F:corrmatrix} Correlation matrices for the lensing signals from the three redshift shells shown in Figure~\ref{F:gglz1z2L123}.  The leftmost panel for z1 reflects higher levels of covariance going to large radial scales.  z1 also has the largest number density of source galaxies, and the overall source number density decreases for z2 and z3.
Going to higher redshifts, a given angular scale corresponds to a larger physical scale and the covariance at large radius moves to the upper right.  At higher redshifts, there are more lens-source pairs within a given subfield due to the angular to physical scale conversion, and the subfield angular scale moves to projected scales much larger than 10 Mpc.  The off-diagonal cells in the panel corresponding to z3 are mainly noise, since most of the pairs come from within a subfield.}
\end{figure}

While the systematics in the data have been addressed and reduced to below the noise level, fits of models to the data must take account of covariances: the errors can be correlated.  The errors in the data shown in Figures~\ref{F:lum} and \ref{F:gglz1z2L123} are correlated (particularly at large scales) due to common sources used in different lens-source pairs that are stacked together. Thus the 13 radial bins are not completely independent in the sense that their errors are correlated.  It is informative to calculate correlation matrices by re-sampling galaxy-galaxy lensing data  for each lens redshift population.  The correlation matrix is the normalized covariance matrix, with matrix elements given by Corr$_{i,j}$ = C$_{i,j}$/(C$_{i,i}$C$_{j,j}$)$^{1/2}$. The covariance matrix C for N jackknife sub-samples is estimated using:
\begin{equation}
C(\Delta \Sigma(R_{i}), \Delta \Sigma(R_{j})) = \frac{N-1}{N}\sum_{k=1}^{N} [\Delta \Sigma ^{k} (R_{i}) - \overline{\Delta \Sigma}(R_{i})] [\Delta \Sigma ^{k} (R_{j}) - \overline{\Delta \Sigma}(R_{j})]
\end{equation}
where $\overline{\Delta \Sigma}(R_{i}) = \sum_{k=1}^{N} \Delta \Sigma^{k} (R_{i})/N$ is the mean value over N sub-samples.  For each of the three lens redshift samples, we take N = 9 random re-samples of all the galaxies in each field. The correlation matrices corresponding to the lensing signals for the three redshift bins in Figure~\ref{F:gglz1z2L123} are shown in Figure~\ref{F:corrmatrix}.  Each cell represents the level of correlation between the given pair of radial bins with darker shades corresponding to higher correlation. There is a higher level of correlation at large radial scales at low lens redshift.  This may be caused by the correlated noise in the dithered 40$^{\prime}$ subfield observing.  At higher redshift 10 Mpc is completely within the subfield angular scale.
Scales of 1 Mpc correspond to angular scales of 4.3$^{\prime}$ for lens sample z1, 2.9$^{\prime}$ for z2, and 2.4$^{\prime}$ for z3.
At the higher redshifts of z2 and z3, there is an increasing number of lens-source pairs within a given 40$^{\prime}$ subfield due to the conversion between angular and physical scales.   However, note that due to our cuts the overall number density of source galaxies decreases going to higher lens redshifts (refer to Table~\ref{tab:lens_src_cuts} for exact numbers).  Thus, the stronger correlations for z1 are also likely linked to the fact that z1 has the largest number density of source galaxies.  While the correlation between bins does not affect the actual values of the galaxy-mass correlation data points, the correlated errors need to be taken into account in fits.  We consider the full correlation matrices in the NFW fits presented in this work.

\end{document}